\documentclass[12pt]{article}
\usepackage{graphicx}
\usepackage{soul}
\usepackage{amsmath, amssymb}
\usepackage{float}
\usepackage[title]{appendix}
\usepackage[hidelinks]{hyperref} 
\usepackage{comment}

\numberwithin{equation}{section} 
\graphicspath{{images/}}
\begin{document}

\bigskip 

\begin{center}
\begin{center}
    {\large \textbf{Mapping Schwarzschild Spacetime: From Kruskal\\to Diamond and Golden Representations}}
\end{center}

\smallskip

\bigskip

Edgar Alejandro León\footnote{ealeon@uas.edu.mx}

\bigskip \

\textit{Facultad de Ciencias Naturales y Exactas, Universidad Autónoma}

\textit{de Sinaloa, 80010, Culiacán, México.}\bigskip

\end{center}
\bigskip \
\bigskip \
\bigskip \ 

\begin{abstract}
\noindent We present several representations of the Schwarzschild spacetime.
Some of these are classical, such as the embedding, Tortoise, and Kruskal
representations. The latter two motivate the introduction of pulsating
coordinates (or the Diamond representation), a compact approach recently
proposed for spherically symmetric static metrics. We then introduce the
Golden representation for the Schwarzschild spacetime, where what we call
the Fibonacci metric, and the golden ratio, become ubiquitous. We conclude
by offering some visual and physical interpretations of this approach.
\end{abstract}

\bigskip \
\bigskip \ 
\bigskip

Keywords: Schwarzschild black hole; Kruskal coordinates; Golden representation

Mathematics Subject Classification 2020: 83A05, 83C15, 83C57

August, 2026

\newpage

\section{Introduction}

\noindent Immediately after Einstein presented the field equations of General Relativity (GR) in 1915, Schwarzschild obtained the first exact
solution to them: the spacetime outside a spherical and non-rotating object. It allowed for the description, to a good approximation, of the geometry
outside astrophysical objects like planets and stars. It was also the prelude for describing the simplest imaginable black hole \cite{MTW}. In the
following decades, alternative descriptions of it were discovered and analyzed, while parallel developments in the realm of cosmology were made 
\cite{Lemaitre}\cite{Tolman}. In black hole theory and cosmology, it became mandatory to explore the subtleties related to the covariance of GR, the
equivalence principle, and the meaning of specific coordinate descriptions in analytic solutions \cite{Bergmann}-\cite{Giovanelli}. The issue of
coordinates and representations continues to be a relevant topic on both fronts \cite{Lemos}-\cite{Visser}.

In this article we analyze different representations of the Schwarzschild spacetime. We start by recalling simple properties that can be extracted
from Schwarzschild coordinates. Then we present some classical representations, such as the near-horizon approximation, the embedding of
spatial sections, and the Tortoise and Kruskal coordinates. We also present two novel representations. The Diamond representation, which was recently
introduced as an extension of a unified scheme for obtaining the Tortoise and Kruskal coordinates \cite{Leon2}. And the \textit{Golden representation}
is introduced here, inspired by previous work highlighting the appearance of the golden ratio in black hole solutions \cite{Nieto1}-\cite{Cruz}. The
transformations involved beautifully map the Schwarzschild spacetime to a fictional spacetime analogous to a relativistic point mass, where the golden
ratio appears ubiquitous.

The rest of the article is structured as follows: In Section \ref{section_2} we present the Schwarzschild solution and briefly review some direct features that can be
extracted from it. Then we obtain the near-horizon approximation alongside its connection with accelerated frames and the Flamm representation. Section
\ref{section_3} presents a derivation of the Tortoise and the Kruskal representations using the same procedure. Then we show the Diamond representation, which
shares a common feature with the former two: they are all (1+1)-conformally flat to hypersurfaces of the Schwarzschild spacetime at constant angles. In
Section \ref{section_4} we introduce the Golden representation for the Schwarzschild spacetime, by slightly deforming the customary Schwarzschild ansatz and
dimensionally reducing the Einstein-Hilbert action, until the Golden coordinates are defined. In Section \ref{section_5} we show how the kinematics of the
(1+1)-dimensional Golden representation yield the same dynamics as the Schwarzschild solution. The arguments of the previous sections allow us to
present different visual Golden representations of the Schwarzschild geometry in Section \ref{section_6}. We end with a summary of the work and some final
comments in Section \ref{section_7}. In the Appendix we present some properties of the Golden ratio and of what we call the \textit{Golden radius}, relating it
with the different representations.

\section{Near Horizon and the Flamm Parabola} \label{section_2}

\noindent Many analytic solutions to the Einstein's field equations, namely%
\begin{equation}
R_{\mu \nu }-\frac{1}{2}g_{\mu \nu }R=8\pi T_{\mu \nu },  \tag{2.1}
\end{equation}%
take the spherically symmetric form \cite{MTW}\cite{Carroll}:

\begin{equation}
ds^{2}=-e^{2\alpha (r)}dt^{2}+e^{2\beta (r)}dr^{2}+r^{2}d\Omega ^{2}, 
\tag{2.2}
\end{equation}%
where $d\Omega ^{2}=d\theta ^{2}+\sin ^{2}\theta d\phi ^{2}$. We use units where $c=G=1$ and Greek indices like $\mu ,\nu $ run from $0$ to $3$, while
Latin indices like $a,b$ run from $1$ to $2$. Many of the solutions share the property $\beta =-\alpha $ \cite{Ayon}-\cite{Leon3}. The interpretation
of a cosmological or black hole origin of the solutions may be subtle. For instance, if the energy-momentum tensor comes from a constant energy-density
filling an isotropic and homogeneous space, the de Sitter or two Lanczos universes can take the same static form. But having distinct spatial
curvature, they take a specific FLRW form in the comoving frame \cite{Leon3}. In a similar vein, assuming a central spherical source can lead to either
the Schwarzschild-de-Sitter or the Schwarzschild solution \cite{Tolman}. The latter is the simplest of all, where $T_{\mu \nu }=0$ in (2.1) outside a
certain radius, implying $R=0$. Einstein's equations reduce to $R_{\mu \nu}=0$, and the metric solution is:%
\begin{equation}
ds^{2}=-\left( 1-\frac{r_{s}}{r}\right) dt^{2}+\frac{dr^{2}}{1-\frac{r_{s}}{r%
}}+r^{2}d\Omega ^{2},  \tag{2.3}
\end{equation}%
where $r_{s}$ is the Schwarzschild radius. This is an appropriate solution outside massive spherical objects, when their angular momentum can be neglected.

A distant observer will observe a time dilation for an event at $r$ given by $dt=\left( -g_{00}\right) ^{-1/2}d\tau$. Then, when a photon travels along a radial geodesic from a given $r$, a faraway observer measures that it is gravitationally redshifted by $\omega
^{\prime }=\omega _{0}\sqrt{-g_{00}}=\omega _{0}\sqrt{1-r_{s}/r}$. We shall refer to $\sqrt{1-r_{s}/r}$ as the redshift factor \cite{Carroll}. Also,
there is a complementary length contraction for the proper radial distance $d\varrho $:

\begin{equation}
dr=d\varrho \sqrt{1-\frac{r_{s}}{r}}.  \tag{2.4}
\end{equation}%
Thus, the $(t,r)$ Schwarzschild coordinates are the time and radial coordinates measured by an observer at infinity. The isometries in Eq. (2.3)
imply the two Killing vectors $T=\partial _{t}$ and $\Phi =\partial _{\phi }$. Spherical symmetry ensures two more independent rotational isometries,
completing a total of four Killing vectors for the Schwarzschild spacetime. Given $T^{\mu }=\delta _{0}^{\mu }$, a static observer has four-velocity $%
u^{\mu }=(-g_{a\beta }T^{\alpha }T^{\beta })^{-1/2}T^{\mu }$. That is, we have $u^{\mu }=\delta _{0}^{\mu }/\sqrt{-g_{00}}$.

Let us approximate the geometry just above $r_{s}$. Since $d\varrho $ in (2.4) represents the proper radial infinitesimal distance at a fixed $r$, we
integrate it:%
\begin{equation}
\varrho =\sqrt{r\left( r-r_{s}\right) }+\frac{r_{s}}{2}\cosh ^{-1}\left( 2%
\frac{r}{r_{s}}-1\right) .  \tag{2.5}
\end{equation}%
For $r>r_{s}$, the identity $\cosh ^{-1}(2r/r_{s}-1)=2\sinh ^{-1}\sqrt{r/r_{s}-1}$ is valid. Then, for $r\cong r_{s}$ one can approximate $\sqrt{%
r\left( r-r_{s}\right) }\approx \sqrt{r_{s}\left( r-r_{s}\right) }$. Since $\sinh ^{-1}z\approx z$ for $z\ll 1$, this yields $\varrho \approx 2r_{s}%
\sqrt{r/r_{s}-1}$, or%
\begin{equation}
r=r_{s}+\frac{\varrho ^{2}}{4r_{s}}.  \tag{2.6}
\end{equation}%
Also, since $\varrho \ll 2r_{s}$, then $1-r_{s}/r=\varrho ^{2}/\left(
\varrho ^{2}+4r_{s}^{2}\right) \approx \varrho ^{2}/(4r_{s}^{2})$. By
rescaling the time as $t=2r_{s}T$, the first two terms in Eq. (2.3) become $%
-\varrho ^{2}dT^{2}+d\varrho ^{2}$.

Consider very small variations in the angle $\theta $ respect to the (arbitrary) $z$-direction. Then $x\cong r_{s}\theta \cos \phi $ and $y\cong
r_{s}\theta \sin \phi $. Thus, $r_{s}^{2}(d\theta ^{2}+\theta ^{2}d\phi ^{2})=dx^{2}+dy^{2}$ can be substituted into the last terms of Eq. (2.3),
resulting \cite{Susskind}:

\begin{equation}
ds^{2}=-\varrho ^{2}dT^{2}+d\varrho ^{2}+dx^{2}+dy^{2}.  \tag{2.7}
\end{equation}%
This approximation is Minkowski spacetime, given the Rindler transformations $t=\varrho \sinh T$ and $z=\varrho \cosh T$. The sections of hyperbolas
induced by $z^{2}-t^{2}=\varrho ^{2}$ when $\varrho =const.$ have slope less than $1$ in a $t-z$ diagram. They are timelike trajectories with constant
acceleration $a=\varrho ^{-1}$. Because Eq. (2.6) implies that $\varrho \rightarrow 0$ as $r\rightarrow r_{s}$, there is a direct
association between the Rindler horizon for a highly accelerated (flat) frame in a radial direction and the Schwarzschild (curved) event horizon 
\cite{Rindler2}\cite{Dray}.

\begin{figure}[H]
    \centering
    \includegraphics[width=0.7\linewidth]{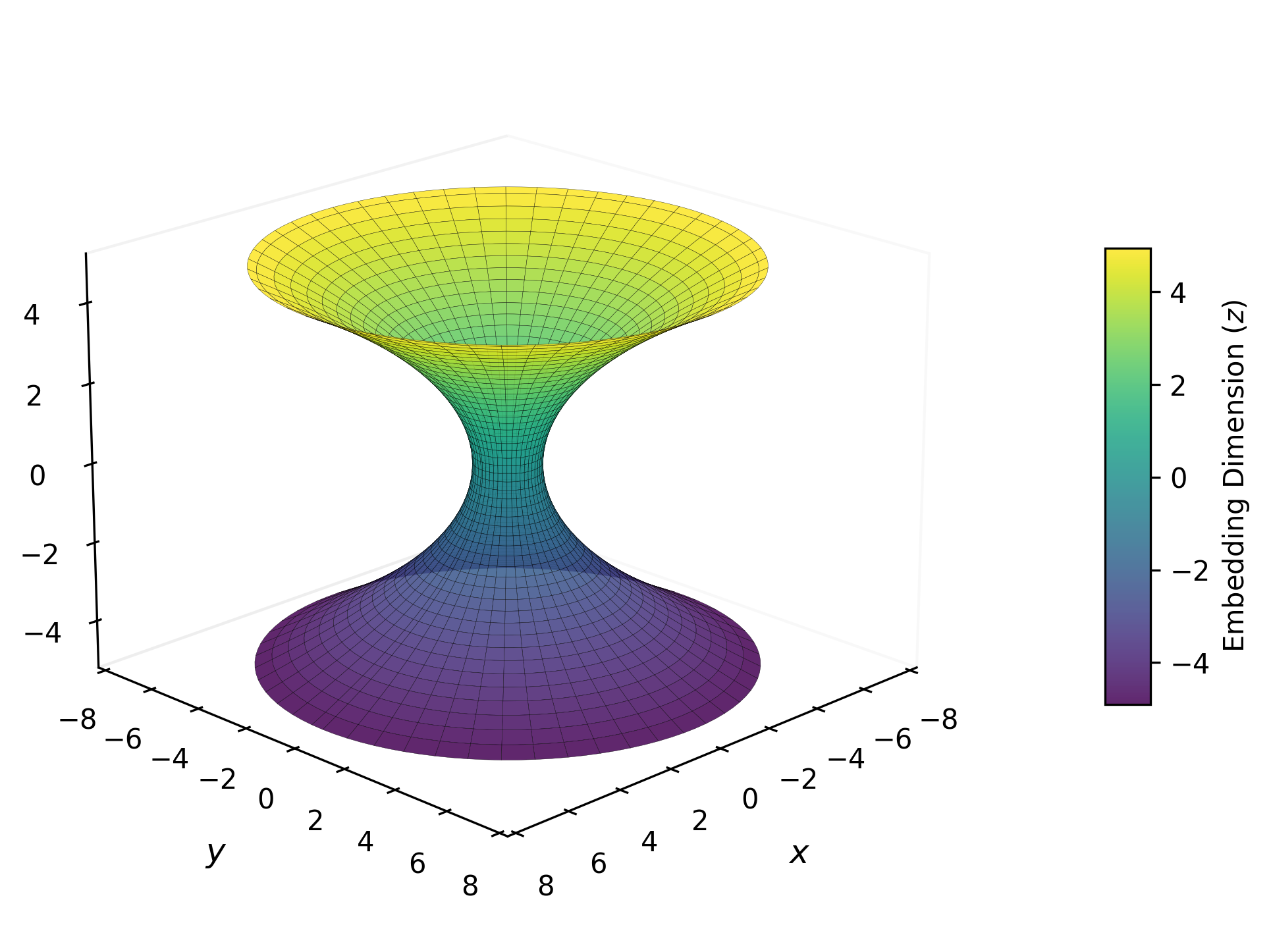}
    \caption{{\protect \footnotesize Flamm Paraboloid. Spatial Embedding of the Schwarzschild Geometry.}}
    \label{Flamm}
\end{figure}

\textit{Embedding diagram}. By setting $dt=d\theta =0$ in Eq. (2.3), we obtain a dimensionally reduced version that can be embedded into a 3D
Euclidean space, for which cylindrical coordinates are appropriate. That is, one embeds the metric $\left( 1-\frac{r_{s}}{r}\right)
^{-1}dr^{2}+r^{2}d\phi ^{2}$ into the target space:%
\begin{equation}
d\Sigma ^{2}=d\rho ^{2}+\rho ^{2}d\phi ^{2}+dz^{2},  \tag{2.8}
\end{equation}%
where $(\rho ,\phi ,z)$ are cylindrical coordinates. By identifying the azimuthal angles, one has $\rho =r$. The two-dimensional surface induces the
dependence:%
\begin{equation}
dz=\sqrt{\frac{r_{s}}{r-r_{s}}}dr.  \tag{2.9}
\end{equation}%
We chose the positive solution, implying a positive slope for the graph $z=z(r)$. This can be integrated to yield $z=\sqrt{4r_{s}\left(
r-r_{s}\right) }$, or%
\begin{equation}
r=\frac{z^{2}}{4r_{s}}+r_{s}.  \tag{2.10}
\end{equation}%
By allowing negative values of $z$ and recalling the azimuthal symmetry in Eq. (2.8), the familiar Flamm paraboloid is obtained (Fig. \ref{Flamm}) \cite%
{GronHervik}.

\noindent This diagram can be associated with two copies of the same spacetime, or with two disjoint spacetimes, with
the circle in $z=0$ acting as the throat of a wormhole connecting them. A special feature of this embedding is that it makes it visually explicit that
traveling along the parabola on a radial path covers more distance than the straight-line Euclidean distance $d\rho $. Given the construction of the
embedding, measuring the radial path along this parabola is equivalent to integrating the proper radial distance in (2.4). That is, Fig. 1 illustrates
the radial length contraction with respect to faraway observers.

Finally, Eqs. (2.6) and (2.10) have the same dependence on $r$: under the near-horizon approximation, the proper radial distance $\varrho $ associated
with Rindler is well approximated by the value $z$ in the Flamm paraboloid.

\section{Kruskal and Pulsating transformations} \label{section_3}

\noindent By setting $d\theta =d\phi =0$ and $ds^{2}=0$ in Eq. (2.3), the local light-cone structure for $r>r_{s}$ is explicitly revealed in $(t,r)$%
-coordinates, in the form:%
\begin{equation}
\frac{dt}{dr}=\pm \frac{dr}{1-\frac{r_{s}}{r}}.  \tag{3.1}
\end{equation}%
In a $t-r$ diagram, this slope is approximately $45{{}^\circ}$ for $r\gg 2M$, but the light cones become highly tilted for $r$ just above
the horizon, with slopes approaching $\pm \infty $ as $r\rightarrow r_{s}$. The Tortoise representation recovers the $\pm 1$ slope \cite{Carroll}. Let
us justify the Tortoise coordinates by using a procedure that allows us to obtain also the Kruskal-Szekeres coordinates \cite{Fronsdal}-\cite{Szekeres}.

We begin by expressing the first two terms on the right-hand side of Eq. (2.3) as $-\gamma dt^{2}+\gamma ^{-1}dr^{2}=\omega ^{2}(-dT^{2}+dR^{2})$,
where $\gamma =1-r_{s}/r$ and $\omega ^{2}$ is the conformal factor. Now consider the metric transformation $g_{ab}=g_{a^{\prime }b^{\prime
}}\partial _{a}x^{a^{\prime }}\partial _{b}x^{b^{\prime }}$, where the indices $a$ and $b$ are $0$ or $1$, and the primed indices refer to the
metric with $\omega ^{2}$. Considering the $00$, $11$ and $01$ components of the transformation, and after rearranging terms, one obtains:%
\begin{equation}
\left( \partial _{t}T\right) ^{2}=\left( \partial _{t}R\right) ^{2}+\omega
^{-2}\gamma ,  \tag{3.2}
\end{equation}%
\begin{equation}
\left( \partial _{r}T\right) ^{2}=\left( \partial _{r}R\right) ^{2}-\omega
^{-2}\gamma ^{-1}  \tag{3.3}
\end{equation}%
and

\begin{equation}
\left( \partial _{t}T\right) ^{2}\left( \partial _{r}T\right) ^{2}=\left(
\partial _{t}R\right) ^{2}\left( \partial _{r}R\right) ^{2}.  \tag{3.4}
\end{equation}%
Inserting Eqs. (3.2) and (3.3) into (3.4) and simplifying yields $\omega^{-2}=\gamma \left( \partial _{r}R\right) ^{2}-\gamma ^{-1}\left( \partial
_{t}R\right) ^{2}$. We can substitute this back into Eq. (3.2) to obtain the simplified form (choosing positive signs when taking the square root):%
\begin{equation}
\partial _{t}T=\gamma \partial _{r}R.  \tag{3.5}
\end{equation}%
Now we assume that $T=T(t)$ and $R=R(r)$. Since the left-hand side of Eq. (5) depends only on $t$ and the right-hand side only on $r$, they must equal
a constant. One may choose $T=t$ in Eq. (3.5), while $\gamma =1-r_{s}/r$ leads to%
\begin{equation}
R=r^{\ast }:=\int \gamma ^{-1}dr=r+r_{s}\ln \left( \frac{r}{r_{s}}-1\right) .
\tag{3.6}
\end{equation}%
The conformal factor then becomes $\omega ^{-2}=\gamma \left( dR/dr\right)^{2}=\gamma ^{-1}$. That is, with the simple dependences $T=T(t)$ and $%
R=R(r) $, we obtain the metric $\gamma (-dT^{2}+dr^{\ast 2})$, described in terms of the Tortoise coordinate $r^{\ast }$ \cite{Possel}. This mapping
pushes the event horizon at $r=r_{s}$ to $r^{\ast }\rightarrow -\infty$. The inversion of Eq. (3.6) is $r=r_{s}\left[ 1+W\left( e^{r^{\ast }/\
r_{s}-1}\right) \right] $, where $W$ is the Lambert W function. Consequently, $r^{\ast }=0$ when $r\approx 1.2786~r_{s}$, and $r^{\ast
}=r_{s}$ when $r=(1+\Omega )r_{s}\approx 1.567~r_{s}$, where $\Omega$ satisfies $\Omega e^{\Omega }=1$ \cite{Valluri}\cite{Gaete}.

An interesting result emerges when applying the same procedure with a (1+1) conformal-Rindler form (instead of Cartesian) \cite{Unruh}\cite{Leon4}. That
is, we now have $-\gamma dt^{2}+\gamma ^{-1}dr^{2}=\omega ^{2}(-\rho^{2}d\tau ^{2}+d\rho ^{2})$. By taking analogous steps to those from Eqs.
(3.2) to (3.5), the analog to Eq. (3.5) is:

\begin{equation}
\rho \partial _{t}\tau =\gamma \partial _{r}\rho .  \tag{3.7}
\end{equation}%
The conformal factor is now $\omega ^{-2}=\gamma \left( \partial _{r}\rho \right) ^{2}-\gamma ^{-1}\left( \partial _{t}\rho \right) ^{2}$. Let us
again assume the simplest functional relationship: $\tau =\tau (t)$ and $\rho =\rho (r)$. Now $d\tau /dt$ is equal to a constant, say $\alpha $,
which allows us to select $\tau =\alpha t$. This implies%
\begin{equation}
\frac{d\rho }{\rho }=\frac{\alpha }{\gamma }dr.  \tag{3.8}
\end{equation}%
Note that the integration involves the Tortoise coordinate from Eq. (3.6), yielding the solution $\rho =\rho _{0}e^{\alpha r^{\ast }}$. This leaves two
integration constants. We set $\rho _{0}=1$, which is the value of $\rho$ when $r^{\ast }=0$, as mentioned earlier. It is convenient to set $\alpha
=1/(2r_{s})$, and then the solution to Eq. (3.8) becomes:%
\begin{equation}
\rho =e^{\frac{r}{2r_{s}}}\sqrt{\frac{r}{r_{s}}-1}.  \tag{3.9}
\end{equation}

The conformal factor now is $\omega ^{-2}=\gamma (d\rho /dr)^{2}$. Performing the derivative yields $\omega ^{2}=4r_{s}^{3}r^{-1}e^{-r/r_{s}}$.
By using the Rindler transformations $\zeta =\rho \sinh \tau $ and $X=\rho \cosh \tau$. Given (3.9) and $\tau =t/2r_{s}$, one obtains:%
\begin{equation}
\begin{array}{c}
\zeta =e^{\frac{r}{2r_{s}}}\sqrt{\frac{r}{r_{s}}-1}\sinh \frac{t}{2r_{s}},
\\ 
\\ 
X=e^{\frac{r}{2r_{s}}}\sqrt{\frac{r}{r_{s}}-1}\cosh \frac{t}{2r_{s}}.%
\end{array}
\tag{3.10}
\end{equation}%
The complete metric in Eq. (2.3) has been transformed to:%
\begin{equation}
ds^{2}=\frac{4r_{s}^{3}e^{-r/r_{s}}}{r}\left( -d\zeta ^{2}+dX^{2}\right)
+r^{2}d\Omega ^{2}.  \tag{3.11}
\end{equation}

\begin{figure}[H]
    \centering
    \includegraphics[width=0.7\linewidth]{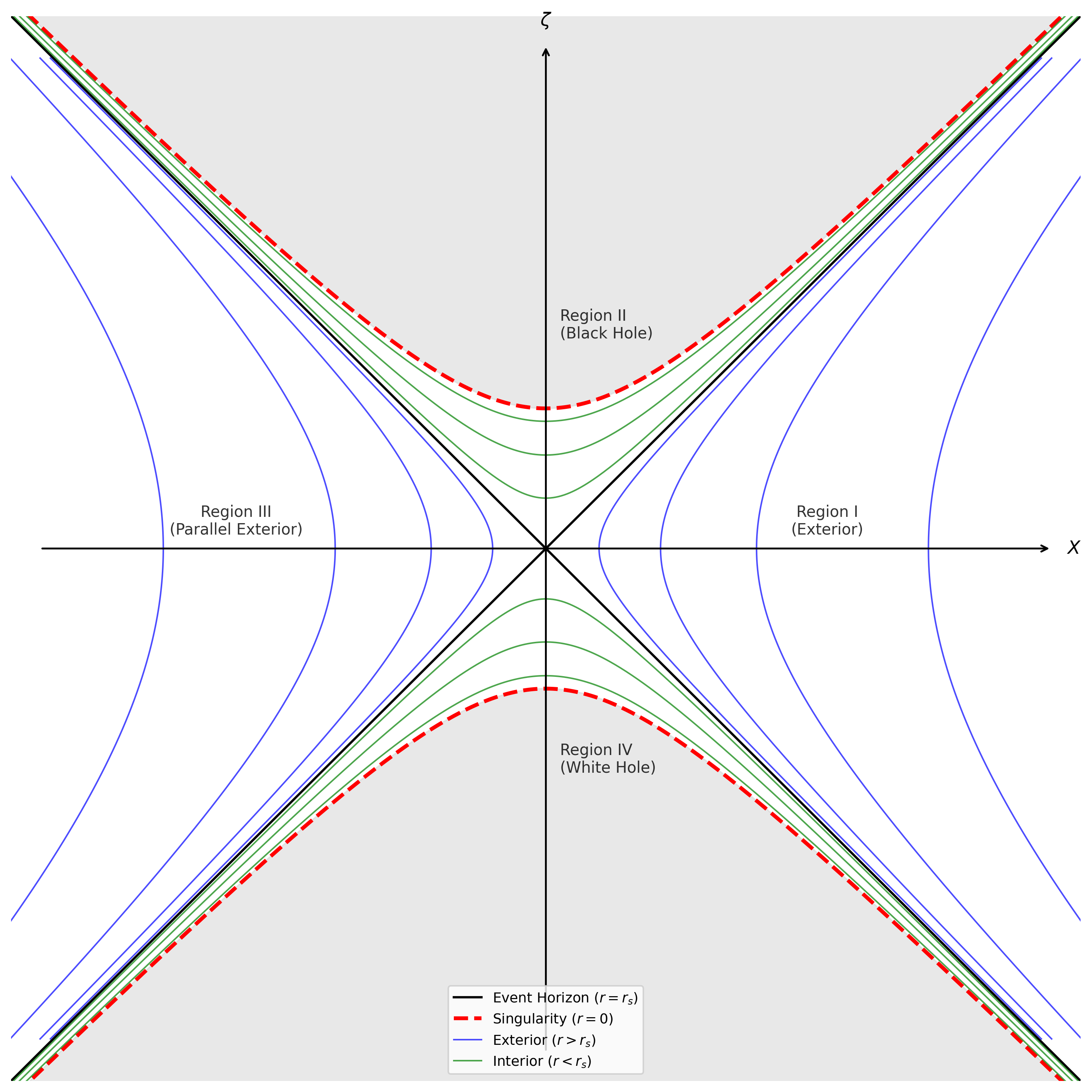}
    \caption{\footnotesize{\protect Maximally extended Schwarzschild spacetime, from the Kruskal representation.}}
    \label{Kruskal}
\end{figure}

Also, from Eq. (3.10) it follows that%
\begin{equation}
X^{2}-\zeta ^{2}=e^{\frac{r}{r_{s}}}\left( \frac{r}{r_{s}}-1\right) , 
\tag{3.12}
\end{equation}%
which relates to the concept of maximal extension. The transformations (3.10) are associated to the exterior region of the black hole (see Fig. \ref{Kruskal}).

\begin{figure}[H]
    \centering
    \includegraphics[width=0.7\linewidth]{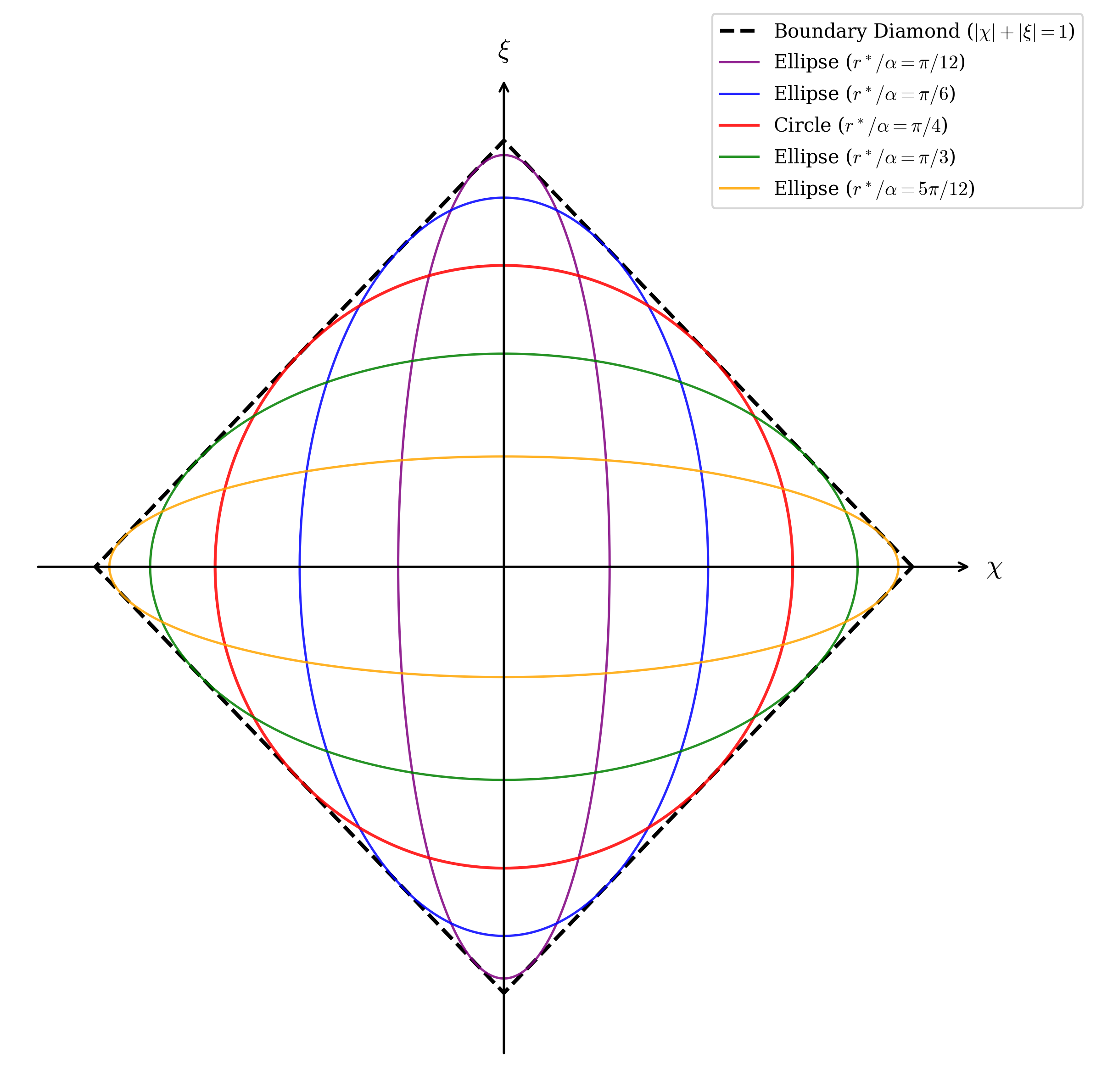}
    \caption{\footnotesize{\protect Compactified Diamond representation of the Schwarzschild spacetime.}}
    \label{Diamond}
\end{figure}

\noindent Note the striking similarities when deriving the Tortoise and Kruskal coordinates. Both describe conformally flat transformations for the
first two terms of Eq. (2.3). There are other possible representations with this feature \cite{Leon2}. In particular, it is possible to obtain a compact representation of the
Schwarzschild spacetime, by means of the \textit{pulsating coordinates} \cite{Leon2}. Instead of hyperbolic trigonometric functions, they are defined completely in terms of trigonometric functions, as:

\begin{equation}
\begin{array}{c}
\xi =D\sin \left( \frac{t}{\alpha }\right) \cos \left( \frac{r^{\ast }}{%
\alpha }\right) , \\ 
\\ 
\chi =D\cos \left( \frac{t}{\alpha }\right) \sin \left( \frac{r^{\ast }}{%
\alpha }\right) ,%
\end{array}
\tag{3.13}
\end{equation}%
where $D$ and $\alpha $ are constants with units of distance. Under these transformations, the metric in Eq. (2.3) becomes equivalent to%
\begin{equation}
ds^{2}=\varpi \left( -d\xi ^{2}+d\chi ^{2}\right) +r^{2}d\Omega ^{2} 
\tag{3.14}
\end{equation}%
where the (1+1) conformal factor is now%
\begin{equation}
\varpi =\frac{2\alpha ^{2}\gamma }{D^{2}}\left[ \cos \left( \frac{2t}{\alpha 
}\right) +\cos \left( \frac{2r^{\ast }}{\alpha }\right) \right] ^{-1}. 
\tag{3.15}
\end{equation}%
From the transformations (3.13), one finds:%
\begin{equation}
\frac{\xi ^{2}}{\cos ^{2}\left( \frac{r^{\ast }}{\alpha }\right) }+\frac{%
\chi ^{2}}{\sin ^{2}\left( \frac{r^{\ast }}{\alpha }\right) }=D^{2}. 
\tag{3.16}
\end{equation}%
Here, we generally have ellipses instead of hyperbolas. The constants $D$ and $\alpha $ govern the amplitude of the representation in a $(\xi ,\chi )$
diagram and the modulation of how rapidly the ellipses change as $r$ varies. In the Kruskal representation there is a similar interpretation: the
amplitude of the representing hyperbolas determines the placement of the upper hyperbola $\zeta ^{2}-X^{2}=1$ in Fig. 2, while the constant $2r_{s}$
inside the hyperbolic trigonometric functions of Eq. (3.10) determines the pace at which the hyperbolas evolve as they approach the event horizon. In
this sense, we set $D=1$ to get the representation in Fig. \ref{Diamond}.

The four corners of the diamond region are located at $(\xi ,\chi )=(\pm1,\pm 1)$. Equation (3.16) allows us to trace changes in the geometry,
starting from a positive value of $r^{\ast }/\alpha $. We choose $\alpha =(2/\pi )(\varphi ^{+}-\ln \varphi ^{+})~r_{s}$, where $\varphi ^{+}=(1+%
\sqrt{5})/2$. This constant is chosen so that when $r^{\ast }/r_{s}=\varphi^{+}-\ln \varphi ^{+}\approx 1.1368$, we have $r^{\ast }/\alpha =\pi /2$,
which yields the horizontal degenerate ellipse (a line segment) at $\xi =0$, with $-1\leq \chi \leq 1$. This somewhat arbitrary initial selection
corresponds to the location $r=\varphi ^{+}r_{s}\approx 1.618$, appearing in the Golden representation discussed below in the following Sections.

\noindent Decreasing $r$ yields horizontal ellipses that become progressively less elongated. When $r^{\ast }/\alpha
=\pi /4$---which in our selection corresponds to the sphere at $r\approx 1.5922~r_{s}$--- the geometry is represented in the diagram by a circle.
Continuing to lower values of $r$ produces vertical ellipses that become increasingly elongated until reaching the vertical degenerate ellipse
bounded by $\xi =\pm 1$ at $r^{\ast }=0$. Continuing to decrease $r$ repeats the inverse process, and this cycle continues infinitely many times as $%
r\rightarrow r_{s}$, since $r^{\ast }\rightarrow -\infty$.

\noindent We have reviewed some classical representations of the Schwarzschild spacetime, such as the Flamm paraboloid and the Tortoise and
Kruskal transformations. We also presented the pulsating coordinates, recently proposed as a natural extension of these classical mappings \cite%
{Leon2}. In turn, the pulsating coordinates have motivated the Causal Diamond coordinates as an alternative to the customary Rindler
representation of Minkowski spacetime \cite{Leon4}. Let us now explore the Golden representation.

\section{Obtaining the Golden Coordinates} \label{section_4}

\noindent Start with a generic spherically symmetric static form, given by:

\begin{equation}
ds^{2}=-e^{2\alpha (r)}dt^{2}+e^{2\beta (r)}dr^{2}+e^{2\gamma (r)}d\Omega
^{2}.  \tag{4.1}
\end{equation}%
In this section we are considering dimensionless units, scaling the coordinates with respect to $r_{s}=1$, for reasons that will be clarified
below.

The Christoffel symbols are calculated from

\begin{equation}
\Gamma _{\mu \nu }^{\alpha }=\frac{1}{2}g^{\alpha \lambda }(\partial _{\mu
}g_{\lambda \nu }+\partial _{\nu }g_{\mu \lambda }-\partial _{\lambda
}g_{\mu \nu }).  \tag{4.2}
\end{equation}%
The non-trivial ones are:

\begin{equation}
\begin{tabular}{lll}
$\Gamma _{01}^{0}=\frac{d\alpha }{dr},$ & $\Gamma _{00}^{1}=e^{2(\alpha
-\beta )}\frac{d\alpha }{dr},$ & $\Gamma _{11}^{1}=\frac{d\beta }{dr},$ \\ 
$\Gamma _{22}^{1}=-e^{2(\gamma -\beta )}\frac{d\gamma }{dr}$ & $\Gamma
_{33}^{1}=-\sin ^{2}\theta e^{2(\gamma -\beta )}\frac{d\gamma }{dr},$ & $%
\Gamma _{12}^{2}=\frac{d\gamma }{dr},$ \\ 
$\Gamma _{33}^{2}=-\sin \theta \cos \theta ,$ & $\Gamma _{13}^{3}=\frac{%
d\gamma }{dr},$ & $\Gamma _{23}^{3}=\cot \theta ,$%
\end{tabular}
\tag{4.3}
\end{equation}%
as well as those related by $\Gamma _{\alpha \beta }^{\mu }=\Gamma _{\beta \alpha }^{\mu }$. Now we substitute them into%
\begin{equation}
R_{\mu \nu }=\partial _{\alpha }\Gamma _{\mu \nu }^{\alpha }-\partial _{\nu
}\Gamma _{\mu \alpha }^{\alpha }+\Gamma _{\alpha \lambda }^{\alpha }\Gamma
_{\mu \nu }^{\lambda }-\Gamma _{\nu \lambda }^{\alpha }\Gamma _{\mu \alpha
}^{\lambda }.  \tag{4.4}
\end{equation}%
The non-trivial Ricci tensor components are then%
\begin{equation}
\begin{array}{c}
R_{00}=e^{2(\alpha -\beta )}\left[ \frac{d^{2}\alpha }{dr^{2}}+\frac{d\alpha 
}{dr}\frac{d(\alpha -\beta +2\gamma )}{dr}\right] , \\ 
R_{11}=\frac{d\alpha }{dr}\frac{d(\beta -\alpha )}{dr}+2\frac{d\gamma }{dr}%
\frac{d(\beta -\gamma )}{dr}-\frac{d^{2}\left( \alpha +2\gamma \right) }{%
dr^{2}}. \\ 
R_{22}=1+e^{2(\gamma -\beta )}\left[ \frac{d(\beta -\alpha -2\gamma )}{dr}%
\frac{d\gamma }{dr}-\frac{d^{2}\gamma }{dr^{2}}\right] , \\ 
R_{33}=\sin ^{2}\theta R_{22}.%
\end{array}
\tag{4.5}
\end{equation}%
Finally, the scalar curvature $R=g^{\mu \nu }R_{\mu \nu }$ results in%
\begin{equation}
R=2e^{-2\beta }\left[ \frac{d(\beta -\alpha )}{dr}\frac{d\left( \alpha
+2\gamma \right) }{dr}-\frac{d^{2}\left( \alpha +2\gamma \right) }{dr^{2}}%
-3\left( \frac{d\gamma }{dr}\right) ^{2}\right] +2e^{-2\gamma }.  \tag{4.6}
\end{equation}

The Einstein-Hilbert action is $16\pi GS=\int \sqrt{-g}Rd^{4}x$. Since $\sqrt{-g}=e^{\alpha +\beta +2\gamma }\sin \theta $, omitting a constant
factor, we are interested in the integral with respect to $r$ of

\begin{equation}
\frac{e^{\alpha -\beta +2\gamma }}{2}\left[ \frac{d(\beta -\alpha )}{dr}%
\frac{d\left( \alpha +2\gamma \right) }{dr}-\frac{d^{2}\left( \alpha
+2\gamma \right) }{dr^{2}}-3\left( \frac{d\gamma }{dr}\right) ^{2}\right] +%
\frac{e^{\alpha +\beta }}{2}.  \tag{4.7}
\end{equation}%
This functional allows us to use the Euler-Lagrange equations. Up to a total
derivative, it is equivalent to the (radial) Lagrangian:

\begin{equation}
L_{r}=\frac{e^{\alpha -\beta +2\gamma }}{2}\left[ 2\frac{d\alpha }{dr}\frac{%
d\gamma }{dr}+\left( \frac{d\gamma }{dr}\right) ^{2}\right] +\frac{e^{\alpha
+\beta }}{2}.  \tag{4.8}
\end{equation}

We define the \textquotedblleft coordinates\textquotedblright \ $q^{1}=\gamma $ and $q^{2}=\alpha $. Then we rewrite $L_{r}$ as%
\begin{equation}
L=\frac{\lambda ^{-1}}{2}\left[ \left( \dot{q}^{1}\right) ^{2}+2\dot{q}^{1}%
\dot{q}^{2}\right] +\frac{\lambda }{2}M^{2},  \tag{4.9}
\end{equation}%
where we have further defined $\lambda ^{-1}=e^{\alpha -\beta +2\gamma }$ and $M^{2}=\lambda ^{-1}e^{\alpha +\beta }=e^{2\alpha +2\gamma }$.
Additionally, for simplicity, we use dot notation for total derivatives with respect to $r$, such as in $\dot{q}^{1}=dq^{1}/dr$. This suggestive form
allows us to treat this Lagrangian as describing a relativistic point particle in two dimensions in the form $\left( \lambda ^{-1}Q_{ab}\dot{q}^{a}%
\dot{q}^{b}+\lambda M^{2}\right) /2$, where $a,b$ run over $\{1,2\}$. We call $Q_{ab}$ the \textit{Fibonacci Q-metric}, with components%
\begin{equation}
Q_{ab}=%
\begin{pmatrix}
1 & 1 \\ 
1 & 0%
\end{pmatrix}%
.  \tag{4.10}
\end{equation}%
We will use $q^{1}$, $q^{2}$ and $\lambda $ as independent variables. The variations with respect to $q^{1}$ and $q^{2}$, with $\partial M^{2}/\partial q^{a}=2M^{2}$, lead to

\begin{equation}
\frac{d}{dr}\left[ \lambda ^{-1}\left( \dot{q}^{1}+\dot{q}^{2}\right) \right]
=\lambda M^{2},  \tag{4.11}
\end{equation}%
and%
\begin{equation}
\frac{d}{dr}\left( \lambda ^{-1}\dot{q}^{1}\right) =\lambda M^{2}, 
\tag{4.12}
\end{equation}%
respectively. The variation with respect to $\lambda $ yields%
\begin{equation}
\lambda ^{-2}\left[ \left( \dot{q}^{1}\right) ^{2}+2\dot{q}^{1}\dot{q}^{2}\right] =M^{2}.  \tag{4.13}
\end{equation}

Here, it is worthwhile to emphasize that these equations are consistent with the ones obtained directly from the original Einstein-Hilbert action \cite%
{Nieto2}. In fact, Eqs. (4.11) and (4.12) together imply:%
\begin{equation}
\frac{d\left( \lambda ^{-1}\dot{q}^{2}\right) }{dr}=0.  \tag{4.14}
\end{equation}%
Recalling the definitions of the coordinates and that $\lambda
^{-1}=e^{\alpha -\beta +2\gamma }$, one can see that, up to a factor, Eq. (4.14) is equivalent to the field equation $R_{00}=0$, which can be read
from (4.5).

Also, by using Eqs. (4.13) and (4.14) in Eq. (4.11), one obtains:%
\begin{equation}
\frac{d}{dr}\left[ \ln \left( \lambda ^{-1}\dot{q}^{1}\right) -q^{1}-2q^{2}%
\right] =0.  \tag{4.15}
\end{equation}%
This is equivalent to $\lambda ^{-1}\dot{q}^{1}=2Ke^{q^{1}+2q^{2}}$, where $K$ is a constant. Returning to the original variables, this is the same as $%
d(e^{\gamma })/dr=Ke^{\alpha +\beta }$. We can use this intermediate result in $\lambda ^{-1}\dot{q}^{2}=const.$ [cf. (4.14)] to show that this implies:%
\begin{equation}
e^{2\alpha }=e^{2q^{2}}=c_{1}-c_{2}e^{-q^{1}},  \tag{4.16}
\end{equation}%
where $c_{1}$ and $c_{2}$ are constants. Again: if one sets $e^{q^{1}}=e^{\gamma }=r$ in the metric (4.1), one recovers the result $\beta
=-\alpha $, with $K=1$. By asymptotic flatness, Eq. (4.16) is the familiar Schwarzschild factor $1-r_{s}/r=-g_{00}$.

\begin{figure}[H]
    \centering
    \includegraphics[width=0.7\linewidth]{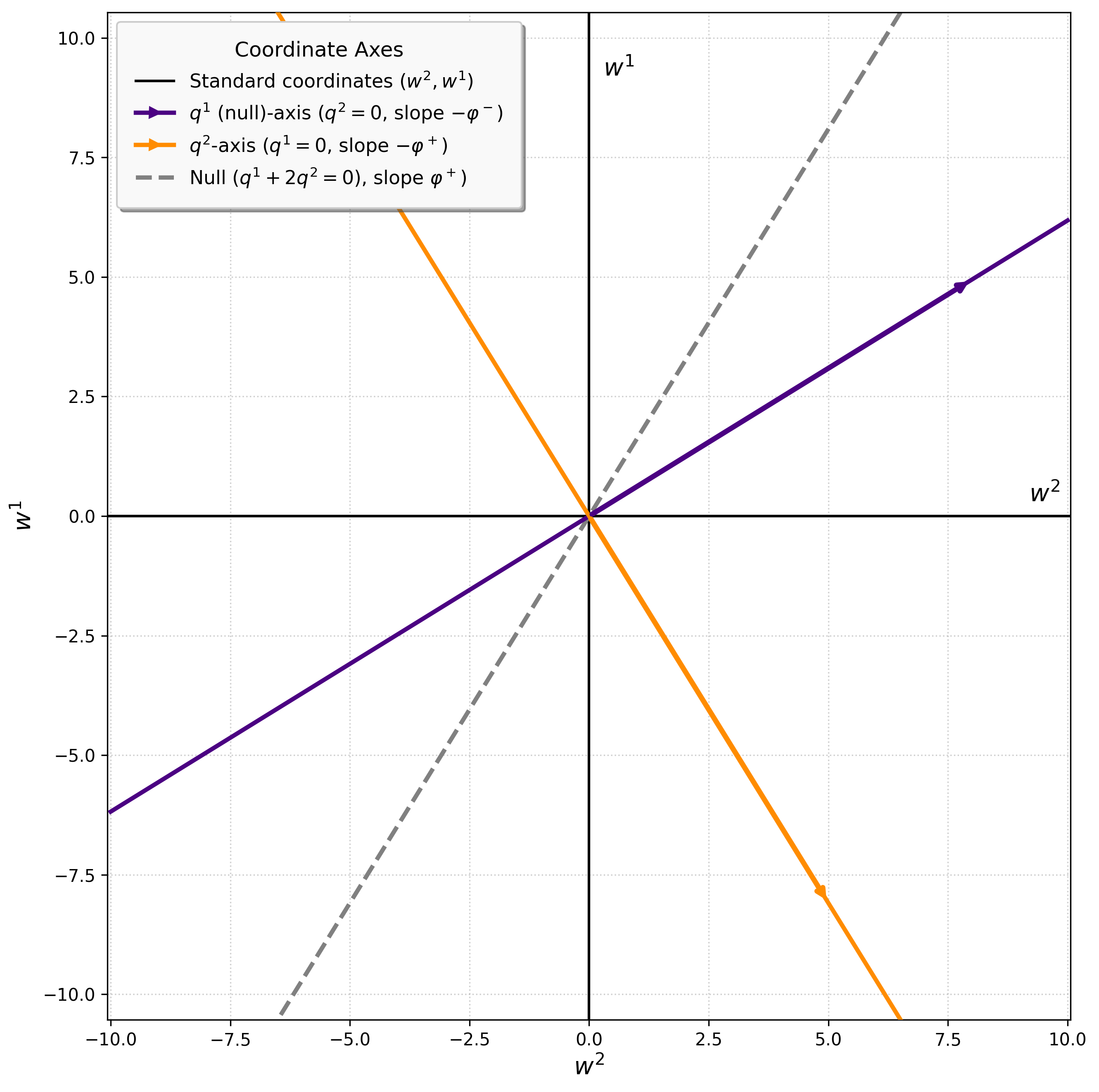}
    \caption{\footnotesize{\protect SO(2) coordinate transformation from $w^{a}$ to $q^{a}$.}}
    \label{SO2}
\end{figure}

We have completed the correspondence with classical results, reaffirming Birkhoff's theorem \cite{Israel}. Now we turn our attention again to the
metric (4.10). Its eigenvalues are $\varphi ^{-}=(1-\sqrt{5})/2$ and $\varphi ^{+}=(1+\sqrt{5})/2$. That is, the negative and positive Golden
ratios have arisen (see the Appendix). By calculating the eigenvectors, one can see that the coordinate transformation%
\begin{equation}
\begin{array}{c}
w^{1}=\frac{1}{5^{1/4}}\left( \sqrt{-\varphi ^{-}}q^{1}-\sqrt{\varphi ^{+}}%
q^{2}\right) , \\ 
\\ 
w^{2}=\frac{1}{5^{1/4}}\left( \sqrt{\varphi ^{+}}q^{1}+\sqrt{-\varphi ^{-}}%
q^{2}\right) ,%
\end{array}
\tag{4.17}
\end{equation}%
implies the equality $Q_{ab}dq^{a}dq^{b}=G_{ab}dw^{a}dw^{b}$, where we have defined:%
\begin{equation}
G_{ab}=%
\begin{pmatrix}
\varphi ^{-} & 0 \\ 
0 & \varphi ^{+}%
\end{pmatrix}%
.  \tag{4.18}
\end{equation}%
Rescaling coordinates in the form $W^{1}=\sqrt{-\varphi ^{-}}w^{1}$, $W^{2}=\sqrt{\varphi ^{+}}w^{2}$, we obtain the usual (1+1)-Minkowski form $\eta
_{ab}dW^{a}dW^{b}$, where $\eta _{ab}=diag(-1,1)$. However, the transformation (4.17) preserves the scale. Its inversion is:%
\begin{equation}
\begin{array}{c}
q^{1}=\frac{1}{5^{1/4}}\left( \sqrt{-\varphi ^{-}}w^{1}+\sqrt{\varphi ^{+}}%
w^{2}\right) , \\ 
q^{2}=\frac{1}{5^{1/4}}\left( -\sqrt{\varphi ^{+}}w^{1}+\sqrt{-\varphi ^{-}}%
w^{2}\right) .%
\end{array}
\tag{4.19}
\end{equation}%
Equations (4.17) and (4.19) are SO(2) transformations. As $w^{1}$ and $w^{2}$ are orthogonal in the induced (1+1)-spacetime, we represent them in a
Minkowski-like diagram. From Eq. (4.19), the $q^{2}$ and $q^{2}$ axes have slopes $\Delta w^{1}/\Delta w^{2}$  equal to $-\varphi ^{-}$ and $-\varphi ^{+}$, respectively. Since $\varphi ^{+}\varphi ^{-}=-1$, they are orthogonal in the diagram, while they are actually not orthogonal in the metric sense. The rotation is by an angle $\arctan \varphi ^{+}\approx 58.28{{}^\circ}$ (see Fig. \ref{SO2}).

From (4.9), (4.10) and (4.18) one has $G_{ab}dw^{a}dw^{b}=Q_{ab}dq^{a}dq^{b}=dq^{1}(dq^{1}+2dq^{2})$. Setting it to zero yields the two null coordinates $q^{1}$ and $q^{1}+2q^{2}$. Equivalently, $G_{ab}dw^{a}dw^{b}=-\left( \sqrt{-\varphi ^{-}} dw^{1}\right) ^{2}+\left( \sqrt{\varphi ^{+}}dw^{2}\right) ^{2}$ admits a direct factorization (see the Appendix). In the diagram, lines $q^{1}=const.$ (as the $q^{2}$ orange axis) have slope of $-\varphi ^{+}$ with respect to the horizontal, while $q^{1}+2q^{2}=const.$ are lines with slope $\varphi ^{+}$, as the grey line.
\noindent For completeness, now we work directly in the Golden representation.

\section{Kinematics in the Golden Representation} \label{section_5}

\noindent So far, we have shown that the dimensional reduction from the (1+3) dynamical Einstein-Hilbert action to a radial equivalent Lagrangian
induces a (1+1)-spacetime. We can now show explicitly that the kinematics in this fiducial spacetime are equivalent to the curvature changes in the
original spacetime. As we have shown, the Lagrangian (4.9) is the same as%
\begin{equation}
L_{r}=\frac{\lambda ^{-1}}{2}Q_{ab}\dot{q}^{a}\dot{q}^{b}+\frac{\lambda }{2}%
M^{2},  \tag{5.1}
\end{equation}%
where $Q_{ab}$ is the Q-Fibonacci metric in Eq. (4.10), while $\lambda^{-1}=e^{-\beta }e^{2q^{1}+q^{2}}$ and $M=e^{q^{1}+q^{2}}$. Recall also the
equality $Q_{ab}\dot{q}^{a}\dot{q}^{b}=G_{ab}\dot{w}^{a}\dot{w}^{b}$, where we have the two equivalent forms (4.10) and (4.18). While $(w^{1},w^{2})$
are explicit orthogonal coordinates in this (1+1) version, $(q^{1},q^{2})$ are the ones with a direct translation in terms of both the Schwarzschild
and the Tortoise coordinates.

Equation (4.13) is equivalent to $Q_{ab}\dot{q}^{a}\dot{q}^{b}=\lambda^{2}M^{2}$, which can also be obtained by varying the action into which
(5.1) is inserted, with respect to $\lambda $. In this (1+1)-dimensional system, the gauge $\lambda =M^{-1}$ would be the equivalent to taking the
normalization $Q_{ab}\dot{q}^{a}\dot{q}^{b}=1$ for the tangent vector $dq^{a}/dr$. However, the connection with the Schwarzschild geometry implies
another norm, as we shall see. Nevertheless, this norm is spacelike, in agreement with all the previous comments about our fictitious particle
system. With all this in mind, we substitute $\lambda =M^{-1}\sqrt{Q_{ab}\dot{q}^{a}\dot{q}^{b}}$ in (5.1), leading to:%
\begin{equation}
L_{r}=M\sqrt{Q_{ab}\dot{q}^{a}\dot{q}^{b}}.  \tag{5.2}
\end{equation}%
This is very similar to the relativistic point particle, but $M$ is not really constant. Since variations of (5.2) lead to the geodesics of this
system, we define the \textit{Golden metric }$\tilde{\Phi}$, incorporating the conformal factor $M^{2}$, as:%
\begin{equation}
d\sigma ^{2}=\Phi _{ab}dq^{a}dq^{b}=M^{2}Q_{ab}dq^{a}dq^{b}.  \tag{5.3}
\end{equation}%
In terms of the rescaled coordinates defined below (4.18), this is explicitly conformal Minkowski, as $M^{2}\eta _{ab}dW^{a}dW^{b}$. The
two-dimensional version of (4.2) for this metric is equivalent to a differential operator acting on $\ln M$ \cite{Carroll}: 
\begin{equation}
\Gamma _{ab}^{c}=(\delta _{b}^{c}\partial _{a}+\delta _{a}^{c}\partial
_{b}-Q^{cd}Q_{ab}\partial _{d})\ln M.  \tag{5.4}
\end{equation}%
Here $\partial _{a}=\partial /\partial q^{a}$ and $Q^{ad}Q_{bd}=\delta_{b}^{a}$ has been used. Given $M=e^{q^{1}+q^{2}}$, one has in general $\ln
M=q^{1}+q^{2}$ and\ $\partial _{a}\ln M=1$. Thus, Eq. (5.4) is the same as $\Gamma _{ab}^{c}=\delta _{a}^{c}+\delta _{b}^{c}-Q^{cd}Q_{ab}\left( \delta
_{d}^{1}+\delta _{d}^{2}\right) $.

The Q-Fibonacci metric (4.10) is $Q_{ab}=\delta _{a}^{1}\delta_{b}^{1}+2\delta _{(a}^{1}\delta _{b)}^{2}=1-\delta _{a}^{2}\delta _{b}^{2}$%
, where parentheses indicate symmetrization as in $A_{(ab)}=\left(A_{ab}+A_{ba}\right) /2$. The inverse Q-metric can be expressed as $%
Q^{cd}=2\delta _{1}^{(c}\delta _{2}^{d)}-\delta _{2}^{c}\delta _{2}^{d}$,implying that $Q^{cd}\left( \delta _{d}^{1}+\delta _{d}^{2}\right) =\delta
_{1}^{c}$. Thus, (5.4) can be expressed as:%
\begin{equation}
\Gamma _{ab}^{c}=\delta _{a}^{c}+\delta _{b}^{c}-\delta _{1}^{c}\left(
1-\delta _{a}^{2}\delta _{b}^{2}\right) .  \tag{5.5}
\end{equation}%
The non-zero Christoffel symbols are $\Gamma _{11}^{1}=\Gamma_{12}^{2}=\Gamma _{21}^{2}=1$ and $\Gamma _{22}^{2}=2$. The geodesic
equation $\ddot{q}^{c}+\Gamma _{ab}^{c}\dot{q}^{a}\dot{q}^{b}=0$ yields:%
\begin{equation}
\ddot{q}^{1}+\left( \dot{q}^{1}\right) ^{2}=0  \tag{5.6}
\end{equation}%
and%
\begin{equation}
\ddot{q}^{2}+2\left( \dot{q}^{1}+\dot{q}^{2}\right) \dot{q}^{2}=0.  \tag{5.7}
\end{equation}

Equation (5.6) can be solved by defining $\zeta =\dot{q}^{1}$, which makes $\ddot{q}^{1}=d\zeta /dr=\dot{q}^{1}(d\zeta /dq^{1})$. One obtains the linear equation $d\zeta /dq^{1}+\zeta=0$. Since the solution is $\dot{q}^{1}=\zeta \propto e^{-q^{1}}$, this in turn implies that the solution to (5.6) is%
\begin{equation}
e^{q^{1}}=Ar+B,  \tag{5.8}
\end{equation}%
where $A$ and $B$ are constants. Let us set $B=0$ and $A=1$, given the angular association and the dimensionless fix $r_{s}=1$ in Eq. (4.1). With
these selections, $e^{q^{1}}=r$ and $\dot{q}^{1}=r^{-1}$. Equation (5.7) can also undergo an order reduction by defining $h=\dot{q}^{2}$, yielding the
equivalent form%
\begin{equation}
\dot{h}+\frac{2}{r}h+2h^{2}=0.  \tag{5.9}
\end{equation}%
With a further definition $h=u^{-1}$, $\dot{h}=-u^{-2}\dot{u}$, Eq. (5.9) is equivalent to $\dot{u}-2r^{-1}u=2$. This is the same as $d\left(
r^{-2}u\right) /dr=2r^{-2}$, which integrates to $u=-2r+Dr^{2}$, with $D$ another integration constant. Since $\dot{q}^{2}=u^{-1}$,%
\begin{equation}
\frac{dq^{2}}{dr}=\frac{1}{r(Dr-2)}=\frac{1}{2}\left( \frac{D}{Dr-2}-\frac{1%
}{r}\right) ,  \tag{5.10}
\end{equation}%
with general solution $q^{2}=\ln \sqrt{D-\frac{2}{r}}+const.$ Setting $D=2$ and $const.=2^{-1}\ln \left( 2^{-1}\right)$, one obtains:%
\begin{equation}
q^{2}=\ln \sqrt{1-\frac{1}{r}}.  \tag{5.11}
\end{equation}

That is, we have recovered $e^{q^{1}}=r$ and also $e^{q^{2}}=\sqrt{1-\frac{1}{r}}$, the redshift factor [see above Eq. (4.9)]. Some translation might be
useful at this point. As we have shown, this representation is in agreement with Birkhoff's theorem, such that the metric (4.1) has as its solution the
explicit Schwarzschild form (2.3). The angular terms in (4.1) imply that (2.3) must be understood in terms of a dimensionless variable $r$. That is,
for the golden representation to work properly, we need to rescale the $r$ coordinate. That is, $r\rightarrow 1$ from above is the same as approaching
the event horizon, which is consistent with selecting $c_{1}=c_{2}=1$ in Eq. (4.16). That is, the selection of constants in the solutions to the
kinematic equations of this (1+1) geometry yields the common Schwarzschild solution, when associating it with the (1+3) version in an appropriate way.
The association is completed if one recalls [below (5.1)] that $\lambda=e^{\beta }e^{-(2q^{1}+q^{2})}$ and $M=e^{q^{1}+q^{2}}$. It follows that the
result $\alpha =q^{2}=-\beta $ is consistent with choosing a gauge where $\lambda =e^{\beta }e^{-q^{1}}/M=e^{-2(q^{1}+q^{2})}=1/M^{2}$.

The association of the kinematics of the Lagrangian (5.2) with the dynamic equations for the metric (4.1) can be summarized as follows:\smallskip

\textit{Golden (1+1) kinematics }$\Longleftrightarrow $\textit{(1+3)-Schwarzschild dynamics.}\smallskip

\noindent Note how the constants have been selected: In (5.8) $B=0$ allows us to have the proper $r^{2}d\Omega ^{2}$ association between (4.1) and
(2.3). $A=1$ allows us to associate in the solutions that $r\rightarrow 1$ is equivalent to approaching the event horizon, and thus one can read $%
r/r_{s}$ every time $r$ appears. Also, selecting $D=2$ and the last constant in (5.11) allows a factoring that also connects the coordinate $q^{2}$ as
(5.11) with the Schwarzschild solution. Then $q^{1}=\ln r$ and $q^{2}=\ln \sqrt{1-1/r}$ are the same as $e^{2\gamma }=r^{2}$ and $e^{2\alpha
}=1-r_{s}/r$, of the Schwarzschild solution, respectively \cite{Carroll}.

We have chosen, from the family of solutions, the following:%
\begin{equation}
e^{2q^{2}}=1-e^{-q^{1}}.  \tag{5.12}
\end{equation}%
This is Eq. (4.16), with $c_{1}=c_{2}=1$. The event horizon $r=1$ has been pushed to $(q^{1},q^{2})\rightarrow (0,-\infty )$, similar to the
Tortoise coordinate behavior.

Although the Golden representation lies in the kinematics induced by the Golden metric (5.3) and its connection with the four-dimensional
Schwarzschild solution, a legitimate question is: what are the intrinsic geometric properties of this fictitious (1+1) spacetime? The short answer
is: it is a flat spacetime. As we saw \ below Eq. (5.5), we have four non-trivial Christoffel symbols for the metric components in (5.3) $\Phi
_{ab}=e^{2(q^{1}+q^{2})}Q_{ab}$, given by $\Gamma _{11}^{1}=\Gamma_{12}^{2}=1$, $\Gamma _{22}^{2}=2$. Since they are constant, the Riemann
tensor for this two-dimensional induced spacetime simplifies to $\tilde{R}_{~~bcd}^{a}=\Gamma _{ce}^{a}\Gamma _{bd}^{e}-\Gamma _{de}^{a}\Gamma
_{bc}^{e}$. By considering the symmetries of the tensor, it turns out that all components are equal to zero, showing that it is flat.

In fact, the explicit transformation to Minkowski form can be seen by writing explicitly the metric (5.3) in terms of the null coordinates
discussed in the last part of Sect. \ref{section_4}, as $d\sigma^{2}=e^{2(q^{1}+q^{2})}dq^{1}(dq^{1}+2dq^{2})$.

Since it can be factored in $dU=e^{q^{1}}dq^{1}$ and $dV:=-e^{(q^{1}+2q^{2})}(dq^{1}+2dq^{2})$, we can make $d\sigma ^{2}=-dUdV$.
Rotating coordinates in the customary way $dV=dT-dX$ and $dU=dT+dX$, yields the Minkowski form $d\sigma ^{2}=-dT^{2}+dX^{2}$. The inversion of
coordinates is $T=e^{q^{1}}(1-e^{2q^{2}})/2$ and $X=e^{q^{1}}(1+e^{2q^{2}})/2 $. By using (5.12) and $e^{q^{1}}=r$, the
Minkowski form of the Golden representation, under translation, is the same as $T=1/2$ and $X=r-1/2$, which in fact can be rescaled and translated. The
special geodesic solution (5.12) which represents the Schwarzschild spacetime, corresponds to a simple \textquotedblleft simultaneity
surface\textquotedblright \ in this Minkowski (1+1) representation, where the space dimension grows linearly with $r$.

We end this section by remarking that there is a third equivalent Lagrangian to (5.1) and (5.2) \cite{Nieto2}. Consider the Legendre transform of (5.1):%
\begin{equation}
H_{A}=\dot{q}^{a}p_{a}-L_{r},  \tag{5.13}
\end{equation}%
where $p_{a}=\partial L_{r}/\partial \dot{q}^{a}$. From (5.1), we obtain $p_{1}=\lambda ^{-1}\left( \dot{q}^{1}+\dot{q}^{2}\right) $ and $%
p_{2}=\lambda ^{-1}\dot{q}^{1}$. We use their inversions to eliminate the $\dot{q}^{a}$-dependence in (5.13):%
\begin{equation}
H_{A}=\frac{\lambda }{2}\left[ \left( 2p_{1}-p_{2}\right) p_{2}-M^{2}\right]
.  \tag{5.14}
\end{equation}%
Now, in terms of the $p^{a}$, the Lagrangian (5.1) satisfies $2L_{r}=\lambda\left( 2p_{1}-p_{2}\right) p_{2}+\lambda M^{2}$. Then (5.14) is equivalent
to $L_{r}=H_{A}-\lambda M^{2}$. Recalling that the inverse to (4.10) is $Q^{cd}=2\delta _{1}^{(c}\delta _{2}^{d)}-\delta _{2}^{c}\delta _{2}^{d}$
[above Eq. (5.5)], we can use (5.13) again to express the compact form for $L_{r}$:%
\begin{equation}
L_{r}=\dot{q}^{a}p_{a}-\frac{\lambda }{2}\left(
p_{a}p_{b}Q^{ab}-M^{2}\right) .  \tag{5.15}
\end{equation}%
This form for the Lagrangian now depends on the momenta $p^{a}$. Nevertheless, it allows for variations with respect to $\lambda $, which
yields the constraint $p_{a}p_{b}Q^{ab}=M^{2}$. By making the translations as before, in particular the identification $\lambda \left(
2p_{1}-p_{2}\right) p_{2}=\lambda ^{-1}Q_{ab}\dot{q}^{a}\dot{q}^{b}$, this Hamiltonian constraint is the same as $Q_{ab}\dot{q}^{a}\dot{q}^{b}=\lambda
^{2}M^{2}$, used for the developments of this Section.

Now let us turn our attention to some visualizations of the Schwarzschild radial geometry, in the $q^{a}$ representation.

\section{Visual Golden Representations} \label{section_6}

\noindent The Golden representation of the Schwarzschild geometry in (2.3) can be understood in a similar way to the Flamm paraboloid in Fig. \ref{Flamm},
defined by the functions $(\rho ,z)=\left( r,\sqrt{4r_{s}\left(r-r_{s}\right) }\right) $. In our present case, the \textit{Golden
coordinates} for the Schwarzschild spacetime are $\left( q^{1},q^{2}\right)=\left( \ln \left( r/r_{s}\right) ,\ln \sqrt{1-r_{s}/r}\right) $. In the
(0+2) Flamm representation of Eq.(2.10), evaluating the interval from $r=r_{s}$ to $r\rightarrow \infty $ corresponds to $(\rho ,z)$ going from $%
(r_{s},0)$ to $(\infty ,\infty )$. In the (1+1) Golden representation, the corresponding limits are from $(q^{1},q^{2})\rightarrow (0,-\infty )$ to $%
(q^{1},q^{2})\rightarrow (\infty ,0)$.

\begin{figure}[H]
    \centering
    \includegraphics[width=0.8\linewidth]{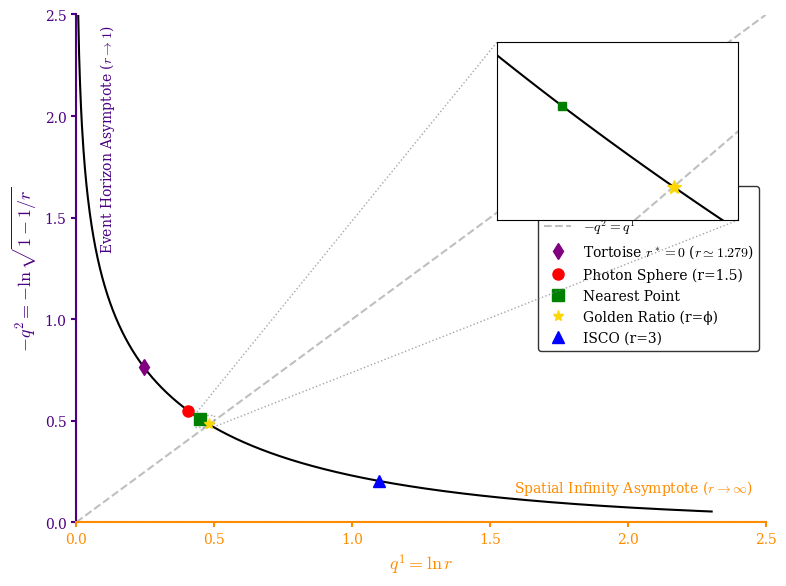}
    \caption{\footnotesize{\protect The Golden curve of Eq. (5.13), showing some relevant points near the \textit{Golden radius}.}}
    \label{Golden-curve}
\end{figure}

Figure \ref{Golden-curve} shows the Golden representation of spatial sections of the Schwarzschild geometry. Note that in Fig. \ref{SO2}, this corresponds to rotating
the coordinates $q^{a}$, followed by an inversion.

\noindent Some important points are explicitly marked, such as the location where $r^{\ast }=0$ ($r\approx 1.2786~r_{s}$), the photon sphere at $r=3r_{s}/2$,
the \textit{Golden radius} ($r=\varphi ^{+}\approx 1.618r_{s}$), and the ISCO (Innermost Stable Circular Orbit) at $r=3r_{s}$. Recall that in the
Diamond representation in Fig. \ref{Diamond} we chose to change vertical ellipses from $r=\varphi ^{+}$ to $r^{\ast }=0$. In Fig. \ref{Golden-curve}, this would represent a change
in the curve from the points marked in yellow and purple.

As we have previously seen, $r=\varphi ^{+}$ is the point where $q^{1}=-q^{2}=\ln \varphi ^{+}$, which corresponds to the intersection of the
curve with the bisector line in this quadrant. However, this is not the closest approach to the point $(0,0)$ in Fig. \ref{Golden-curve}. While the distance to the
curve (indicating $r=\varphi ^{+}$) is $\sqrt{2}\ln \varphi ^{+}\approx0.68054$, the shortest distance (minimizing and numerically solving) is $%
d\approx 0.67830$. By Eq. (4.17), this point has w-coordinates $(w^{1},w^{2})=\frac{1}{5^{1/4}}\left( \varphi ^{+}\right) ^{-3/2}\ln \varphi
^{+}\left[ \left( \varphi ^{+}\right) ^{3},1\right] $. In Fig. \ref{SO2}, the line joining $(0,0)$ to the point representing $r=\varphi ^{+}$ has slope $\left(
\varphi ^{+}\right) ^{3}$. This line forms an angle of $\pi /2-(1/2)\arctan(1/2)$ with the horizontal, approximately $76.72{{}^\circ}$. There, the family of possible solutions to (4.16) lies in the region where $q^{1}>0$, $q^{2}<0$. There we can set $c_{1}=1$ by the asymptotic flatness of the Schwarzschild solution, while the constant $c_{2}$ modulates $r$ with respect to some radius.

By means of (4.17), the translation to the w-representation is as follows: taking from $r$ from $r_{s}$ to $\infty $ corresponds to changing from $%
(w^{1},w^{2})\rightarrow (\infty ,-\infty )$ to $(w^{1},w^{2})\rightarrow(\infty ,\infty )$. For $r\gg 1$, they behave as $w^{1}\simeq 5^{-1/4}\sqrt{%
-\varphi ^{-}}\ln r$ and $w^{2}\simeq 5^{-1/4}\sqrt{\varphi ^{+}}\ln r$, and we have the asymptotic behavior $w^{1}=-\varphi ^{-}w^{2}$. This corresponds
to the line $q^{1}$ in Fig. \ref{SO2}. On the other hand, the near-horizon approximation $r\approx 1$ has the asymptotic behavior $w^{1}\simeq -\varphi
^{+}w^{2}$, with $w^{1}\simeq -5^{-1/4}\sqrt{\varphi ^{+}}q^{2}$ and $w^{2}\simeq 5^{-1/4}\sqrt{-\varphi ^{-}}q^{2}$.

\begin{figure}[H]
    \centering
    \includegraphics[width=0.8\linewidth]{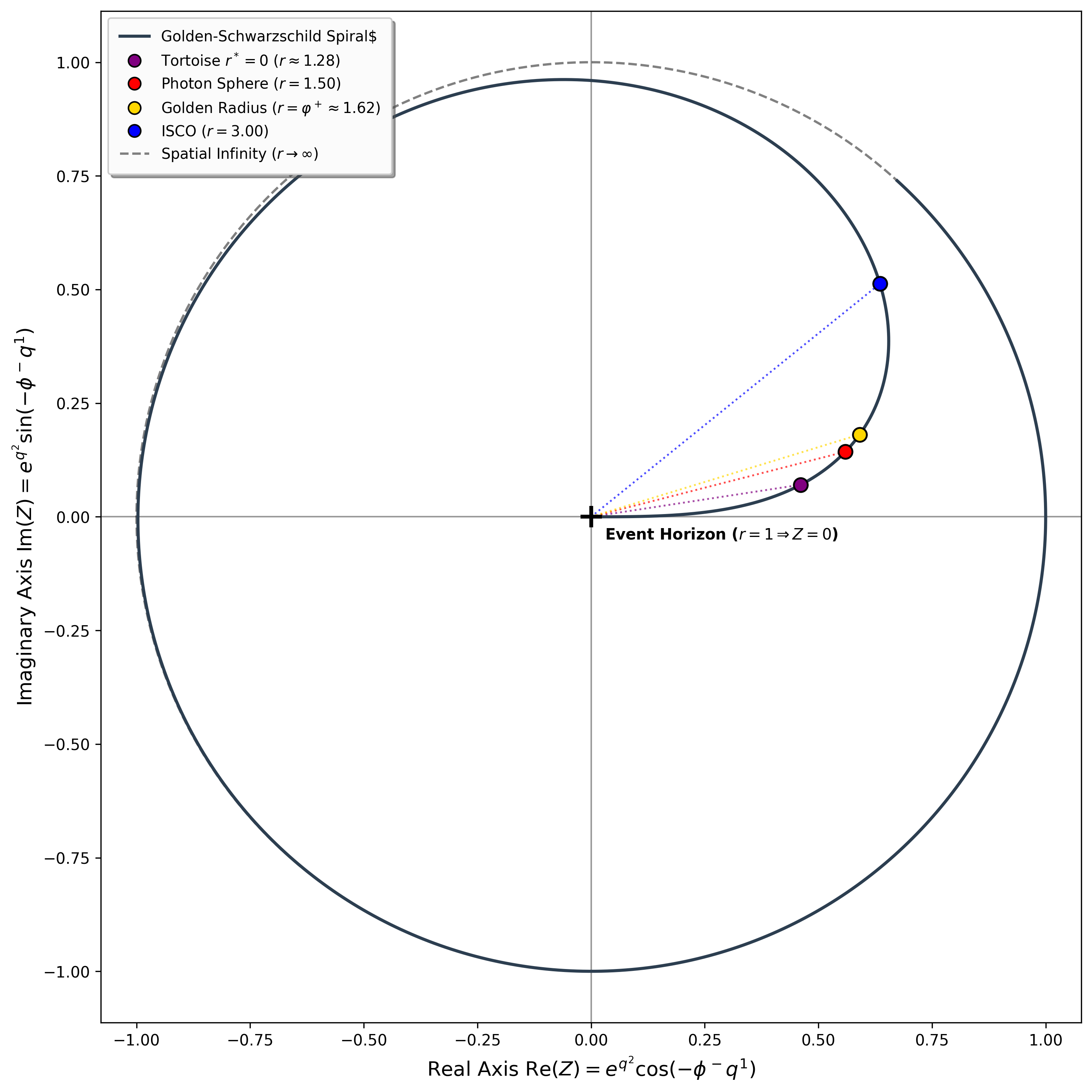}
    \caption{\footnotesize{\protect The complex Golden Schwarzschild Spiral. Spatial infinity has been compactified to $\left
\vert Z\right \vert =1$, while the event horizon is at the origin.}}
    \label{Golden-spiral}
\end{figure}

We now use the $w^{a}$ coordinates to get an alternative representation. Let us first rewrite the transformations (4.17) as:%
\begin{equation}
\begin{array}{c}
w^{1}=\cos \alpha ~q^{1}-\sin \alpha ~q^{2}, \\ 
\\ 
w^{2}=\sin \alpha ~q^{1}+\cos \alpha ~q^{2},%
\end{array}
\tag{6.1}
\end{equation}%
where $\cos \alpha =5^{-1/4}\sqrt{-\varphi ^{-}}$ and $\sin \alpha =5^{-1/4}\sqrt{\varphi ^{+}}$, and the rotation angle is $\arctan \varphi ^{+}$.
Recall also that, with $W^{1}=\sqrt{-\varphi ^{-}}w^{1}$ and $W^{2}=\sqrt{\varphi ^{+}}w^{2}$, one has that $G_{ab}dw^{a}dw^{b}=\eta _{ab}dW^{1}dW^{2}$%
, where $G_{ab}=diag(\varphi ^{-},\varphi ^{+})$ and $\eta _{ab}=diag(-1,1)$. This motivates performing a Wick rotation for the timelike coordinate $%
W^{1}$ as $W^{1}\rightarrow iT$ \cite{Gibbons}. This transformation would allow us to consider a complex representation $W^{2}+iW^{1}$, but in fact in
the complex plane this would be equivalent to Fig. \ref{Golden-curve}. Recalling that (5.13) is such that it preserves the scale between the systems $w^{a}$ and $%
q^{a}$, this instead motivates the introduction of the complex quantity $\sqrt{\varphi ^{+}}q^{2}+i\sqrt{-\varphi ^{-}}q^{1}$. Factoring out $\sqrt{%
\varphi ^{+}}$, we simply exponentiate $q^{2}-i\varphi ^{-}q^{1}$, obtaining:

\begin{equation}
Z=e^{q^{2}}e^{-i\varphi ^{-}q^{1}}.  \tag{6.2}
\end{equation}%
The properties of this construction, represented in Fig. \ref{Golden-spiral}, are: the horizon has been mapped to the origin $Z=0$, while the curve asymptotically
approaches the unit circle as $r\rightarrow \infty $. The role of the null coordinate $q^{1}$ is to modulate the oscillations of the solution, as in
the imaginary periodic time induced by a Wick rotation. The line segment joining the origin to a given point measures directly the redshift factor $%
\sqrt{1-1/r}=e^{q^{2}}$, while the angle $\theta =-\varphi ^{-}q^{1}$ allows us to interpret $q^{1}=\ln r$ as a periodic variable.

\noindent The previous representative points of Fig. \ref{Golden-curve} are shown in the Golden-Schwarzschild spiral of Fig. \ref{Golden-spiral}: the radius where $r^{\ast }=0$, the
Golden radius $r=\varphi ^{+}$, the photon sphere $r=1.5$, and the ISCO at $r=3$. Consider again the point where $r=\varphi ^{+}$: its distance to the
origin is $\left \vert Z\right \vert =\sqrt{1-1/\varphi ^{+}}=-\varphi ^{-}$. Also, the squared distance of this point to the unit circle is $%
1-\left\vert Z\right \vert ^{2}=1-\left( \varphi ^{-}\right) ^{2}=-\varphi^{-}$. Thus, the point representing $r=\varphi ^{+}$ has a distance $%
-\varphi ^{-}\approx 0.618$ from the origin, while it is separated by $\sqrt{-\varphi ^{-}}\approx 0.786$ from the unit circle. Interestingly, the
squared distance to the point at infinity in the complex representation can also have a spiral representation, given that $1-\left \vert Z\right \vert
^{2}=r^{-1}=e^{-\varphi ^{+}\theta }$, where we used (5.12) in $\left \vert Z\right \vert ^{2}=e^{2q^{2}}$.

Notice that this diagram allows us to easily spot points where $r$ is of order unity. The first crossing of the vertical axis occurs when $\varphi
^{+}\pi /2=q^{1}$, corresponding to $r=\exp (\varphi ^{+}\pi /2)\approx 12.7$. On the left, the first crossing with the horizontal axis is when $r=\exp
(\varphi ^{+}\pi )\approx 161.3$, while the second crossing with the vertical axis is below, when $r=\exp (3\varphi ^{+}\pi /2)\approx 2048$.

Finally, to obtain a real representation, consider that Eq. (6.2) is $Z=\left \vert Z\right \vert e^{i\theta }$, with $\left \vert Z\right \vert
=e^{q^{2}}$ and $\theta =-\varphi ^{-}q^{1}$. Given that $q^{1}=\ln r$, the inversion is%
\begin{equation}
r=e^{\varphi ^{+}\theta }.  \tag{6.3}
\end{equation}%
This describes a real logarithmic spiral \cite{Anatriello}.

\section{Final comments: Linking representations} \label{section_7}

\noindent In this article, we have discussed several representations of the Schwarzschild spacetime: the near-horizon approximation, the Flamm
paraboloid, as well as the Tortoise, Kruskal, Diamond, and Golden representations. While the former are classical representations and the
Diamond representation is a recent development, the Golden representation is newly introduced in this article.

We have delved into the relationships between these representations. For instance, the Flamm parabola allows us to read the integration of (2.4), the
radial proper distance for the Schwarzschild metric, through the Euclidean distance traveled on the parabola (2.10). From the Flamm
parabola one can also measure the $z$ distance for $r\cong r_{s}$ to obtain $\varrho$, the radial Rindler approximation (2.7) of the near-horizon approximation.

We also highlighted a derivation of the Tortoise and Kruskal coordinates from the unified scheme of considering conformal (1+1) radial sections
in the Schwarzschild geometry. The Tortoise coordinate (3.6) appears when considering the Cartesian form, while the Kruskal coordinates appear naturally by doing the analogous developments for a conformal flat Rindler transformation. We then presented the Diamond representation, the other main general possibility for these types of transformations \cite{Leon2}.

In the last three sections we introduced the Golden representation, arising from transformations of the Einstein-Hilbert action. The radial
Lagrangian results in a description similar to the kinematical Lagrangian of a relativistic particle \cite{Nieto1}\cite{Nieto2}. In this article, we have
deepened the significance of this dimensional reduction, clarified issues regarding the correct translation with Schwarzschild coordinates, and
proposed useful visualizations.

Let us make three final remarks about the representations we have explored. The first point is that the Golden radius $r=\varphi ^{+}r_{s}$ unites all
of them in an interesting sense (see the Appendix). This radius corresponds to:

\begin{itemize}
\item The location on the Flamm parabola $(\varrho ,z)=(\varphi ^{+},2\sqrt{%
-\varphi ^{-}})r_{s}$.

\item One of infinitely many radii that yield a horizontal degenerate ellipse in the Diamond representation (Fig. \ref{Diamond}), given the $\alpha $ chosen.

\item A Tortoise coordinate $r^{\ast }/r_{s}=\varphi ^{+}-\ln \varphi ^{+}$.

\item The point $q^{1}=-q^{2}=\ln \varphi ^{+}$ of the Golden representation (Fig. \ref{Golden-curve}).

\item A location $Z=-\varphi ^{-}e^{-i\varphi ^{-}\ln \varphi ^{+}}$ in the complex representation (Fig. \ref{Golden-spiral}).
\end{itemize}

\noindent Furthermore, it is the radius where the proper distance (2.10) and the Tortoise coordinate (3.6) satisfy:

\begin{equation}
3r^{\ast }+2\varrho =\left( \varphi ^{+}\right) ^{4}r_{s}.  \tag{7.1}
\end{equation}%
This equation beautifully links the Flamm and Tortoise representations, and mimics the purely numeric relation $3\varphi ^{+}+2=\left( \varphi
^{+}\right) ^{4}$.

A second point worth mentioning is the relation of the Golden representation to the others presented. For instance, the Golden representation retains a
mixture of properties from the embedding and the Tortoise representations. Just as the Euclidean distance in the Flamm parabola allows one to read the
integrated proper distance $\varrho $, the complex Golden representation (Fig. \ref{Golden-spiral}) allows one to directly read the redshift factor $\sqrt{1-r_{s}/r}$ by measuring the distance to the origin (the event horizon). Also, the direct $q^{1}-q^{2}$ diagram of Fig. \ref{Golden-curve} implies that the event horizon has been pushed to $-\infty $ in $q^{2}$, just as it is in the Tortoise coordinate (3.6).

Furthermore, while the first null coordinate is $q^{1}=\ln (r/r_{s})$, the second one is, in the Schwarzschild interpretation (see the Appendix):

\begin{equation}
q^{1}+2q^{2}=\frac{r^{\ast }-r}{r_{s}}.  \tag{7.2}
\end{equation}%
This measures how much the Tortoise advances past the Schwarzschild coordinate. Recalling from Eq. (3.9) that $\rho ^{2}=e^{\frac{r^{\ast }}{%
r_{s}}}$, where $\rho $ is the intermediate Rindler-type coordinate leading to the Kruskal transformations, exponentiating (7.2) yields the Kruskal
connection $e^{q^{1}+2q^{2}}e^{\frac{r}{r_{s}}}=\rho ^{2}$.

There is also a relation to the Diamond representation, beyond just choosing a specific $\alpha $ in Eq. (3.16). By taking any $\alpha $ in Eqs.
(3.13)-(3.16), one can see that the periodicity in the change of ellipses in Fig. \ref{Diamond} is driven by the argument $r^{\ast }/\alpha $, while the
oscillations in the spiral representations proposed at the end of Section \ref{section_6} are driven by the angle $\theta =-\varphi ^{-}q^{1}$. However, while both
make infinite oscillations when $r\rightarrow \infty $, only the Diamond representation makes infinite oscillations when approaching the event
horizon. The relation (6.2), together with (7.2) and the translations $q^{1}=\ln (r/r_{s})$ and $e^{2q^{2}}=1-r_{s}/r=\left \vert Z\right \vert
^{2} $, allows us to obtain the relationship:%
\begin{equation}
\frac{r^{\ast }}{\alpha }=\frac{r_{s}}{\alpha }\left( \varphi ^{+}\theta
+\ln \left \vert Z\right \vert ^{2}+e^{\varphi ^{+}\theta }\right) . 
\tag{7.3}
\end{equation}%
This equation highlights two asymptotic connections. For $r\rightarrow r_{s}$, which has $\theta \rightarrow 0$ in the golden spirals, the logarithm
dominates, and the near-horizon pulsations are governed by $r^{\ast }/\alpha\cong (r_{s}/\alpha )\ln \left \vert Z\right \vert ^{2}$. On the other
extreme, as $r\rightarrow \infty $, the exponential dominates, yielding $r^{\ast }/\alpha \cong (r_{s}/\alpha )e^{\varphi ^{+}\theta }$.

The third and final point concerns other physical connections of the Golden representation and possible future work. For instance, notice that
integrating $2(dr^{\ast }-dr)$ yields what one would call a \textit{radial Shapiro time delay} for a two-way light path, by setting the impact
parameter to zero \cite{MTW}\cite{Shapiro}. That is, by simply taking the difference of the null coordinate (7.2) at two distinct points, one obtains
the time delay effect: $\Delta t_{delay}=2r_{s}\Delta (q^{1}+2q^{2})$. Other physical aspects, such as thermodynamic aspects of the black hole, the use
of Golden transformations or extensions as a toy arena to discuss implementations from a lower-dimensional manifold to four-dimensional or
extradimensional ones, and the possible use of the phase space version of the Golden representation for deforming the two-dimensional flat version to
obtain curved spacetimes, are left for future work (see Refs. \cite{Bergmann2}-\cite{Camelia} for interesting possibilities regarding these
issues).

\section*{Acknowledgments}
The author thanks J. A. Nieto and A. Meza for useful comments on this subject. He also recognizes the \textit{Secretar%
\'{\i}a de Ciencia, Humanidades, Tecnolog\'{\i}a e Innovaci\'{o}n (SECIHTI)} for support through the \textit{Sistema Nacional de Investigadoras e Investigadores
(SNII)}.

\appendix

\section*{Appendix: Golden ratio, and Golden radius}

\noindent The golden ratio arises from the standard continuous division condition $\varphi =(\varphi -1)^{-1}$, yielding the defining relation \cite%
{Basin}\cite{Basin2}:%
\begin{equation}
\varphi ^{2}=\varphi +1.  \tag{A.1}
\end{equation}%
Solving the quadratic leads to two values: $\varphi ^{+}=\left( 1+\sqrt{5}\right) /2$ and $\varphi ^{-}=\left( 1-\sqrt{5}\right) /2$. From this, we
have $\varphi ^{+}\varphi ^{-}=-1$, $\varphi ^{+}+\varphi ^{-}=1$, and $\varphi ^{+}-\varphi ^{-}=\sqrt{5}$. By successive multiplications of (A.1)
by $\varphi $, one can note that the following relation holds for integers $n\geq 0$:%
\begin{equation}
\left( \varphi ^{\pm }\right) ^{n+2}=\left( \varphi ^{\pm }\right)
^{n+1}+\left( \varphi ^{\pm }\right) ^{n}.  \tag{A.2}
\end{equation}%
Thus, $\left( \varphi ^{\pm }\right) ^{2}=\left( 3\pm \sqrt{5}\right) /2$, $\left( \varphi ^{\pm }\right) ^{3}=\left( 4\pm 2\sqrt{5}\right) /2$ and $%
\left( \varphi ^{\pm }\right) ^{4}=\left( 7\pm 3\sqrt{5}\right) /2$. The Fibonacci sequence, starting with $F_{0}=0$, has elements $%
(0,1,1,2,3,5,8...) $. By starting with (A.1) and calculating integer powers of\ $\left( \varphi ^{\pm }\right) ^{n}$, one can convert them to a linear
form in $\varphi ^{\pm }$:%
\begin{equation}
\left( \varphi ^{\pm }\right) ^{n}=F_{n}\varphi ^{\pm }+F_{n-1}.  \tag{A.3}
\end{equation}%
In the relations $\sqrt{\varphi ^{+}}+\sqrt{-\varphi ^{-}}$ and $\sqrt{\varphi ^{+}}-\sqrt{-\varphi ^{-}}$, we can factor out $\sqrt{\varphi ^{+}}$
and $\sqrt{-\varphi ^{-}}$, respectively. By considering again that $\varphi^{+}\varphi ^{-}=-1$ and $\varphi ^{+}=1-\varphi ^{-}$, one obtains:%
\begin{equation}
\sqrt{\varphi ^{+}}+\sqrt{-\varphi ^{-}}=\varphi ^{+}\sqrt{\varphi ^{+}}, 
\tag{A.4}
\end{equation}%
and

\begin{equation}
\sqrt{\varphi ^{+}}-\sqrt{-\varphi ^{-}}=-\varphi ^{-}\sqrt{-\varphi ^{-}}. 
\tag{A.5}
\end{equation}%
\bigskip Given the transformations (4.19), using (A.4) and (A.5), and making similar factorizations as before, the null coordinate $q^{1}+2q^{2}$ is%
\begin{equation}
q^{1}+2q^{2}=\frac{1}{\sqrt{5}}\left[ -\sqrt{-\varphi ^{-}}w^{1}+\sqrt{\varphi ^{+}}w^{2}\right] ,  \tag{A.6}
\end{equation}%
while the other null coordinate is $q^{1}=\frac{1}{5^{1/4}}\left( \sqrt{-\varphi ^{-}}w^{1}+\sqrt{\varphi ^{+}}w^{2}\right) $.

Now we address some properties of the Golden radius $r=\varphi ^{+}r_{s}$. Consider the proper distance integrated from $r=\varphi ^{+}r_{s}$ to $%
r=r_{s}$, which corresponds to measuring the distance over the the Flamm parabola (Fig. 1), from $z=0$ to the point representing the Golden radius.
We rewrite Eq. (2.5) as:

\begin{equation}
\varrho =\sqrt{r\left( r-r_{s}\right) }+r_{s}\ln \left( \sqrt{\frac{r}{r_{s}}%
}+\sqrt{\frac{r}{r_{s}}-1}\right) ,  \tag{A.7}
\end{equation}%
where we used the identity $\cosh ^{-1}u=\ln (u+\sqrt{u^{2}-1})$. At $r=r_{s} $, $\varrho =0$, while for $r=\varphi ^{+}r_{s}$, one has $\varrho
/r_{s}=\sqrt{-\varphi ^{+}\varphi ^{-}}+\ln \left( \sqrt{\varphi ^{+}}+\sqrt{-\varphi ^{-}}\right) $. By $\varphi ^{+}\varphi ^{-}=-1$ and (A.5), one has 
$\varrho /r_{s}=1+(3/2)\ln \left( \varphi ^{+}\right) $. Also, by evaluating Eq. (2.10) at $r=\varphi ^{+}r_{s}$ yields $z=r_{s}\sqrt{4\left( \varphi
^{+}-1\right) }$. Thus the location of the Golden point on the Flamm parabola is where $z=2\sqrt{-\varphi ^{-}}r_{s}$.

From the identification $q^{1}=\ln (r/r_{s})$ and $q^{2}=\ln \sqrt{1-r_{s}/r}$, it follows that the null coordinate $q^{1}+2q^{2}$ is the same as $\ln
\left( r/r_{s}-1\right) $. That is, by considering Eq. (3.6), we have the simple relation:%
\begin{equation}
q^{1}+2q^{2}=\frac{r^{\ast }-r}{r_{s}}.  \tag{A.8}
\end{equation}%
It ranges over $(-\infty ,\infty )$, being zero when $\ln \left(r/r_{s}-1\right) =0$, where $r=2r_{s}$. It is positive when $r>2r_{s}$ and
negative for $r\in (r_{s},2r_{s})$. Precisely in this region lies the Golden radius, where $q^{1}+2q^{2}=-\ln \varphi ^{+}$.

From (3.7) one has $r^{\ast }/r_{s}=\varphi ^{+}-\ln \varphi ^{+}$. Although (A.7) is not generally related to the Tortoise coordinate, at the Golden
radius $\varrho /r_{s}=1+(3/2)\left( \varphi ^{+}-r^{\ast }/r_{s}\right) $. Verifying the properties of $\varphi ^{+}$ below (A.2), at $r=\varphi
^{+}r_{s}$, we have:%
\begin{equation}
3r^{\ast }+2\varrho =\left( \varphi ^{+}\right) ^{4}r_{s}.  \tag{A.9}
\end{equation}%
This beautiful equation relates proper distance, Tortoise coordinate and the Golden ratio in the Schwarzschild geometry, at $r=\varphi ^{+}r_{s}$. The
approximated values there are $\varrho \approx 1.7218r_{s}$ (radial length contraction), $r^{\ast }\approx 1.1368r_{s}$, and $\left( \varphi
^{+}\right) ^{4}\approx 6.8541$. Note the striking similarity of $3r^{\ast}+2\varrho =\left( \varphi ^{+}\right) ^{4}r_{s}$ with $3\varphi
^{+}+2=\left( \varphi ^{+}\right) ^{4}$, obtained from (A.4).

In the Golden spiral defined by Eq. (6.2), $r=\varphi ^{+}r_{s}$ has the complex representation $Z=-\varphi ^{-}e^{-i\varphi ^{-}\ln \varphi ^{+}}$%
.

\end{document}